\documentclass[apl,reprint,twocolumn,superscriptaddress]{revtex4}

\usepackage{amsmath, amsfonts, amssymb}
\usepackage{graphicx}
\usepackage{epstopdf}

\begin{document}

\title{Ultrafast all-optical gated amplifier based on ZnO nanowire lasing}

\author{Marijn A. M. Versteegh}
\affiliation{Debye Institute for Nanomaterials Science, Utrecht University, Princetonplein 1, 3584 CC Utrecht, The Netherlands}
\affiliation{Kavli Institute of Nanoscience, Delft University of Technology, Lorentzweg 1, 2628 CJ Delft, The Netherlands}
\author{Peter J. S. van Capel}
\affiliation{Debye Institute for Nanomaterials Science, Utrecht University, Princetonplein 1, 3584 CC Utrecht, The Netherlands}
\author{Jaap I. Dijkhuis\footnote{j.i.dijkhuis@uu.nl}}
\affiliation{Debye Institute for Nanomaterials Science, Utrecht University, Princetonplein 1, 3584 CC Utrecht, The Netherlands}

\begin{abstract}
We present an ultrafast all-optical gated amplifier, or transistor, consisting of a forest of ZnO nanowire lasers. A gate light pulse creates a dense electron-hole plasma and
excites laser action inside the nanowires. Source light traversing the nanolaser forest is amplified, partly as it is guided through the nanowires, and partly as it propagates
diffusively through the forest. We have measured transmission increases at the drain up to a factor 34 for 385-nm light. Time-resolved amplification measurements show that the
lasing is rapidly self-quenching, yielding pulse responses as short as 1.2 ps.
\end{abstract}

\maketitle All-optical computing is potentially much faster than conventional electronic computing. For the development of ultrafast all-optical computing, ultrafast
all-optical transistors and logic gates are required. However, nonlinear optical effects on which ultrafast all-optical components must be based are almost invariably very
weak in conventional materials, thereby limiting possible applications. Solutions are therefore pursued in specially designed materials, such as plasmonic nanorod
metamaterials \cite{wurtz 2011} and periodically poled lithium niobate crystals \cite{zhang 2011}, where these nonlinear effects are greatly enhanced.

Lasing in ZnO nanowires forms an interesting opportunity in this context. ZnO nanowires have been shown to exhibit strong laser action between about 385 nm and 390 nm under
optical excitation \cite{huang 2001, johnson 2001, johnson 2003, van vugt 2006}. This laser action requires gain lengths of a few micrometers only. Indeed, modal gain lengths
in this range have recently been measured inside single ZnO nanowires \cite{richters 2012}. Time-resolved measurements on lasing ZnO nanowires have been performed using
several techniques: optical injection probing \cite{szarko 2005, song 2005}, Kerr gating \cite{kwok 2005, mitsubori 2009}, sum-frequency gating \cite{song 2008}, and by using
a streak camera \cite{fallert 2008, thonke 2009, xing 2011}. All reports show that under strong excitation the laser action lasts very short: its duration can be shorter than
2 ps. This short lasing time contrasts with the luminescence lifetimes of tens or hundreds of picoseconds that were measured below the laser threshold.

Here we present an ultrafast all-optical gated amplifier consisting of a forest of ZnO nanowire lasers. Its operation at room temperature is analyzed by time-resolved
amplification measurements. A source light pulse is strongly amplified by a forest of ZnO nanowire lasers if it arrives shortly after an excitation gating pulse. We have
measured an on-off ratio of 34 at the drain and a pulse response as short as 1.2 ps.

\begin{figure}
\begin{center}
\includegraphics{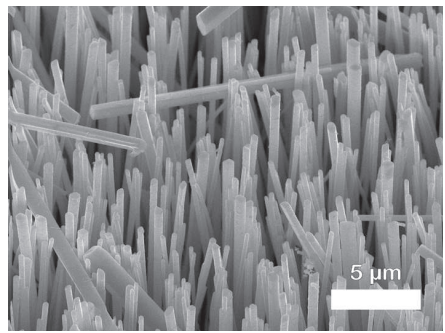}
\caption{SEM image of our ultrafast all-optical gated amplifier: a forest of 20 $\mu$m long ZnO nanowire lasers. \label{6figBSG}}
\end{center}
\end{figure}

Coupling a light pulse into a single nanowire is cumbersome and suffers from severe losses. In order to obtain robust amplification we have chosen to use a dense forest of ZnO
nanowires as gated amplifier. Nanowire forests exhibit exceptionally small reflectivity losses, so virtually all input light propagates into the forest \cite{hu 2007, muskens
2008, zhu 2009, kupec 2010}. Using a nanowire forest as gated amplifier also has the virtue that the source light strongly scatters at the nanowires and the seed film. The
resulting diffusive motion of the source photons through the forest increases the average time photons spend inside the sample from values around 100 fs to values around 1 ps
\cite{versteegh 2012}. The increased length over which gain takes place critically enhances the total amplification of the transistor.

A scanning-electron-microscope (SEM) image of our transistor is shown in Fig. \ref{6figBSG}. This is the same nanowire forest as used for earlier light diffusion measurements
\cite{versteegh 2012}. The nanowires were epitaxially grown on a sapphire substrate, using the carbothermal method \cite{prasanth 2006}. First a porous ZnO seed film was
created. On top of that, the nanowires were grown, with their crystal $c$-axes parallel to the wires. The wires of our transistor are all about 20 $\mu$m long. Their diameters
vary between 100 and 500 nm with an average of 250 nm. We measured the nanowire density to be 0.85 $\mu$m$^{-2}$, and the ZnO filling fraction 0.08.

\begin{figure}
\begin{center}
\includegraphics[width=0.47\textwidth]{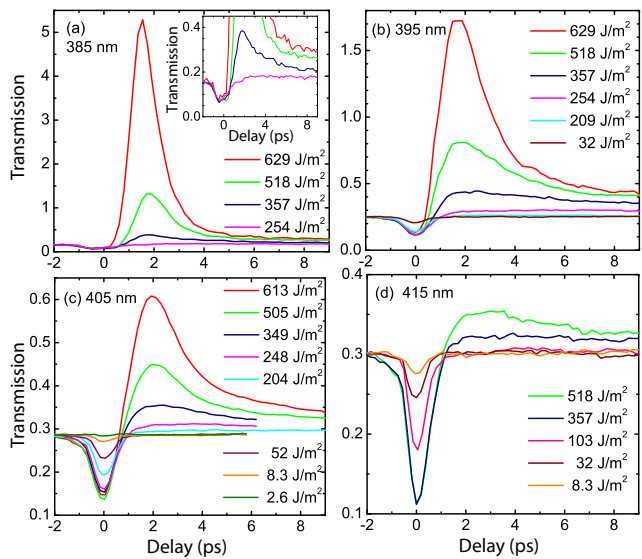}
\caption{(Color online) Time-resolved amplification experiment on the ZnO nanowire forest shown in Fig. 1: Measured transmission of a source pulse vs delay with respect to the 800-nm excitation gate pulse for (a) 385
nm, (b) 395 nm, (c) 405 nm, and (d) 415 nm source wavelength, and specified gate fluences.\label{6figgain}}
\end{center}
\end{figure}

The nanowire forest transistor is gated by a 125-fs, 800-nm pulse from an amplified Ti:sapphire ``Hurricane" laser. Since 800-nm photons have an energy of 1.55 eV and the band
gap of ZnO is 3.37 eV at room-temperature, 800-nm absorption in ZnO is a three-photon process \cite{he 2005, dai 2005, gu 2008, versteegh prb}. At high intensities this
three-photon excitation leads to the formation of a dense electron-hole plasma, whereby optical gain and possibly laser action take place \cite{versteegh prb, zhang 2006, dai
2011, versteegh nanowire lasing}. The advantage of using three-photon excitation is that the nanowire forest can be excited over its entire thickness in an approximately
homogeneous way. Alternatively, gating by ultraviolet above-band-gap pulses is possible, but the small penetration depth of ultraviolet light (50 nm) in ZnO limits the
fraction of the system that is excited, and thereby the amplification that is obtained.

In our experiment, a source pulse traverses the nanowire forest with a tunable delay with respect to the 800-nm gating pulse. As source pulses we used 125-fs pulses of four
wavelengths: 385 nm, 395 nm, 405 nm, and 415 nm, each with 2-nm bandwidth. These pulses were created via white-light generation in a sapphire crystal and subsequent
sum-frequency generation by mixing this white light with 800-nm light in a beta barium borate crystal, allowing frequency selection by angle tuning \cite{versteegh ol}. If the
source pulse arrives at the nanowire forest transistor after it has been gated by the excitation pulse, the stored electronic energy is drained and the source light is
amplified. This amplification occurs partly as source light is guided through the excited wires, and partly as scattered source light diffuses through the excited forest.
Transmitted source light is collected by a lens and measured by a photodiode and a lock-in amplifier. To calibrate our results, we measured the fraction of transmitted light
that is collected by the lens. We divided our measurement results by that fraction to get the total transmission.

Results of the ultrafast signal response of our optical transistor for all four source wavelengths are presented in Fig. \ref{6figgain}. For gate fluences above 200 J/m$^2$
the transistor opens up. We observe increased signal transmission when the source pulse arrives at the nanowire forest after the gate pulse. The amplification of the signal
rapidly increases with gate fluence and we observe a strong wavelength dependence. At 629 J/m$^2$ we observe for 395-nm source light a factor 7 increase in transmission, from
0.25 to 1.7. For 385 nm an on-off ratio as high as $5.3/0.15=34$ is found.

Fast decay of the amplification becomes visible for fluences above 300 J/m$^2$. The response becomes faster with increasing gate fluence. At 629 J/m$^2$ the duration of the
385-nm amplification is only 1.2 ps (full width at half maximum). The remarkable strength of the observed amplification, in combination with its very short duration, makes
this ZnO nanowire forest especially suitable as an ultrafast optical UV amplifier.

For zero delay the transmission of the source pulse is reduced (Fig. \ref{6figgain}). This is caused by two-photon absorption of a source photon and a gate photon,
simultaneously present inside the sample. This effect forms the mechanism of our ultrafast bulk ZnO all-optical shutter \cite{versteegh ol}. Inside a ZnO nanowire forest the
same phenomenon can be used to measure the photon diffusion \cite{versteegh 2012}.

\begin{figure}
\begin{center}
\includegraphics{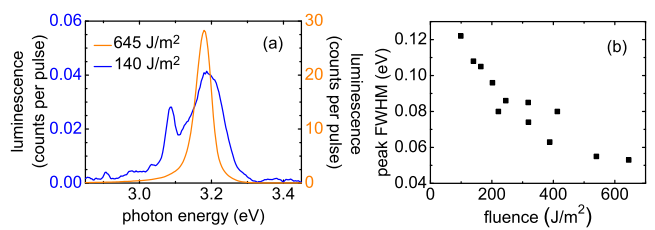}
\caption{Time-integrated photoluminescence experiment on a similar ZnO forest under 800-nm excitation. (a) Measured emission spectra, showing an electron-hole recombination peak centered at 3.2 eV and an second harmonic peak centered at 3.1 eV. The scale on the left corresponds to the data at 145 J/m$^2$. The scale on the right corresponds to the data at 645 J/m$^2$. (b) Width (FWHM) of the electron-hole recombination peak vs excitation fluence.}
\end{center}
\end{figure}

The fast decay of amplification we observe for high gate fluences signifies rapid decay of charge carriers. This fast carrier decay can be ascribed to laser action. The
amplification in our experiment (Fig. \ref{6figgain}) is so strong that lasing must occur in the nanowire forest, mediated by the electron-hole plasma. This interpretation is
confirmed by the emission spectra shown in Fig. 3, which were measured on a similar ZnO nanowire forest. The sharpening of the emission peak for increasing excitation
intensity agrees with a gradual transition from spontaneous emission to lasing.

Studies on the quantum efficiency of ZnO nanowires \cite{zhang 2005, gargas 2011} show that below the laser threshold the external luminescence quantum efficiency is at most
10-20\%, which means that most of the charge carriers decay nonradiatively. Above the laser threshold, however, the external quantum efficiency rises to 60\%, and the internal
quantum efficiency even to 85\% \cite{zhang 2005}. Apparently, the laser action is so efficient that carrier decay by lasing dominates nonradiative losses, including nonlinear
effects as Auger recombination. This increasing quantum efficiency is also observed in our experiments, where we found that the time-integrated luminescence intensity due to
electron-hole recombination (so excluding the second harmonic peak at 3.1 eV) increases more strongly than proportional to $F^{3}$, where $F$ is the excitation fluence. We
conclude that at high excitations there is a self-quenching gain: strong stimulated emission causes the majority of the carriers to rapidly recombine radiatively, leading to
an ultrafast reduction of the gain. The gain quenching is assisted by the fact that the reflectivities at the wire ends are small \cite{maslov 2003, bordo 2008}.
Self-quenching gain also explains the short lasing times in ZnO nanostructures reported in Refs. \cite{szarko 2005, song 2005, kwok 2005, song 2008, mitsubori 2009, fallert
2008, thonke 2009, xing 2011}.

In conclusion, we have demonstrated that a forest of ZnO nanowire lasers acts as an ultrafast all-optical gated amplifier. Three-photon absorption of an 800-nm gating pulse
leads to a net amplification of an incident 385-nm source pulse up to a factor 5, and an on-off ratio of 34 at the drain. The strong optical gain and lasing in the nanowire
forest are rapidly self-quenching. The amplification time can be as short as 1.2 ps. The gating fluences needed are 300-600 J/m$^2$, low enough to allow gating of the fast
transistor by high-repetition-rate mode-locked lasers. This nanowire all-optical UV transistor may have applications in optical computing and can be used in ultrafast
pump-probe experiments. Nanowire forests of other direct semiconductors should permit nanowire optical transistors at other wavelengths.

We thank Henrik Porte, Benjamin Brenny, Wouter Ensing, Ruben van der Wel, and Bas Zegers for contributing to the experiments, Dani\"{e}l Vanmaekelbergh and Heng-Yu Li for
providing the nanowire forests, and Cees de Kok and Paul Jurrius for technical assistance.


\begin{thebibliography}{99}
\bibitem{wurtz 2011}
G. A. Wurtz, R. Pollard, W. Hendren, G. P. Wiederrecht, D. J. Gosztola, V. A. Podolskiy, and A. V. Zayats, Nature Nanotech. \textbf{6,} 107 (2011).
\bibitem{zhang 2011}
Y. Zhang, Y. Chen, and X. Chen, Appl. Phys. Lett. \textbf{99,} 161117 (2011).
\bibitem{huang 2001}
M. H. Huang, S. Mao, H. Feick, H. Yan, Y. Wu, H. Kind, E. Weber, R. Russo, and P. Yang, Science \textbf{292,} 1897 (2001).
\bibitem{johnson 2001}
J. C. Johnson, H. Yan, R. D. Schaller, L. H. Haber, R. J. Saykally, and P. Yang, J. Phys. Chem. B \textbf{105,} 11387 (2001).
\bibitem{johnson 2003}
J. C. Johnson, H. Yan, P. Yang, and R. J. Saykally, J. Phys. Chem. B \textbf{107,} 8816 (2003).
\bibitem{van vugt 2006}
L. K. van Vugt, S. R\"{u}hle, and D. Vanmaekelbergh, Nano Lett. \textbf{6,} 2707 (2006).
\bibitem{richters 2012}
J. P. Richters, J. Kalden, M. Gnauck, C. Ronning, C. P. Dietrich, H. von Wenckstern, M. Grundmann, J. Gutowski, and T. Voss, Semicond. Sci. Technol. \textbf{27,} 015005
(2012).
\bibitem{szarko 2005}
J. M. Szarko, J. K. Song, C. W. Blackledge, I. Swart, S. R. Leone, S. Li, and Y. Zhao, Chem. Phys. Lett. \textbf{404,} 171 (2005).
\bibitem{song 2005}
J. K. Song, J. M. Szarko, S. R. Leone, S. Li, and Y. Zhao, J. Phys. Chem. B \textbf{109,} 15749 (2005).
\bibitem{kwok 2005}
W. M. Kwok, A. B. Djuri\u{s}i\'{c}, Y. H. Leung, W. K. Chan, and D. L. Phillips, Appl. Phys. Lett. \textbf{87,} 093108 (2005).
\bibitem{mitsubori 2009}
S. Mitsubori, I. Katayama, S. H. Lee, T. Yao, and J. Takeda, J. Phys.: Condens. Matter \textbf{21,} 064211 (2009).
\bibitem{song 2008}
J. K. Song, U. Willer, J. M. Szarko, S. R. Leone, S. Li, and Y. Zhao, J. Phys. Chem. C \textbf{112,} 1679 (2008).
\bibitem{fallert 2008}
J. Fallert, F. Stelzl, H. Zhou, A. Reiser, K. Thonke, R. Sauer, C. Klingshirn, and H. Kalt, Opt. Express \textbf{16,} 1125 (2008).
\bibitem{thonke 2009}
K. Thonke, A. Reiser, M. Schirra, M. Feneberg, G. M. Prinz, T. R\"{o}der, R. Sauer, J. Fallert, F. Stelzl, H. Kalt, S. Gsell, M. Schreck, and B. Stritzker, Adv. Solid. State Phys. \textbf{48,} 39 (2009).
\bibitem{xing 2011}
G. Z. Xing, D. D. Wang, B. Yao, A. Q. Lloyd Foong Nien, and Y. S. Yan, Chem. Phys. Lett. \textbf{515,} 132 (2011).
\bibitem{hu 2007}
L. Hu and G. Chen, Nano Lett. \textbf{7,} 3249 (2007).
\bibitem{muskens 2008}
O. L. Muskens, J. G\'{o}mez Rivas, R. E. Algra, E. P. A. M. Bakkers, and A. Lagendijk, Nano Lett. \textbf{8,} 2638 (2008).
\bibitem{zhu 2009}
J. Zhu, Z. Yu, G. F. Burkhard, C. M. Hsu, S. T. Connor, Y. Xu, Q. Wang, M. McGehee, S. Fan, and Y. Cui, Nano Lett. \textbf{9,} 279 (2009).
\bibitem{kupec 2010}
J. Kupec, R. L. Stoop, and B. Witzigmann, Opt. Express \textbf{18,} 27589 (2010).
\bibitem{versteegh 2012}
M. A. M. Versteegh, R. E. C. van der Wel, and J. I. Dijkhuis, Appl. Phys. Lett. \textbf{100,} 101108 (2012).
\bibitem{prasanth 2006}
R. Prasanth, L. K. van Vugt, D. A. M. Vanmaekelbergh, and H. C. Gerritsen, Appl. Phys. Lett. \textbf{88,} 181501 (2006).
\bibitem{dai 2005}
D. C. Dai, S. J. Xu, S. L. Shi, M. H. Xie, and C. M. Che, Opt. Lett. \textbf{30,} 3377 (2005).
\bibitem{he 2005}
J. He, Y. Qu, H. Li, J. Mi, and W. Ji, Opt. Express \textbf{13,} 9235 (2005).
\bibitem{gu 2008}
B. Gu, J. He, W. Ji, and H. T. Wang, J. Appl. Phys. \textbf{103,} 073105 (2008).
\bibitem{versteegh prb}
M. A. M. Versteegh, T. Kuis, H. T. C. Stoof, and J. I. Dijkhuis, Phys. Rev. B \textbf{84,} 035207 (2011).
\bibitem{zhang 2006}
C. F. Zhang, Z. W. Dong, G. J. You, S. X. Qian, and H. Deng, Opt. Lett. \textbf{31,} 3345 (2006).
\bibitem{dai 2011}
J. Dai, C. X. Xu, Z. L. Shi, R. Ding, J. Y. Guo, Z. H. Li, B. X. Gu, and P. Wu, Opt. Mater. \textbf{33,} 288 (2011).
\bibitem{versteegh nanowire lasing}
M. A. M. Versteegh, D. Vanmaekelbergh, and J. I. Dijkhuis, Phys. Rev. Lett. \textbf{108,} 157402 (2012).
\bibitem{versteegh ol}
M. A. M. Versteegh and J.I. Dijkhuis, Opt. Lett. \textbf{36,} 2776 (2011).
\bibitem{zhang 2005}
Y. Zhang, R. E. Russo, and S. S. Mao, Appl. Phys. Lett. \textbf{87,} 043106 (2005).
\bibitem{gargas 2011}
D. J. Gargas, H. Gao, H. Wang, and P. Yang, Nano Lett. \textbf{11,} 3792 (2011).
\bibitem{maslov 2003}
A. V. Maslov and C. Z. Ning, Appl. Phys. Lett. \textbf{83,} 1237 (2003).
\bibitem{bordo 2008}
V. G. Bordo, Phys. Rev. B \textbf{78,} 085318 (2008).
\end{thebibliography}
\end{document}